\documentclass[preprint,aps,showpacs]{revtex4}

\usepackage[dvips]{graphicx}

\usepackage{dcolumn}     % Align table columns on decimal point

\renewcommand{\case}{\frac}

\begin{document}
 
{

\title{Quadratic momentum dependence in the nucleon-nucleon interaction}

\author{R. B. Wiringa\cite{rbw}}

\affiliation{Physics Division, Argonne National Laboratory, Argonne, IL 60439}

\author{A. Arriaga\cite{aa}}

\affiliation{Centro de Fisica Nuclear da Universidade de Lisboa, Avenida
Gama Pinto 2, 1699 Lisboa, Portugal \\ 
and Departamento de Fisica, Faculdade de Ci\^{e}ncias da Universidade de 
Lisboa, 1700 Lisboa, Portugal}

\author{V. R. Pandharipande\cite{vrp}}

\affiliation{Department of Physics, University of Illinois at Urbana-Champaign,
1110 West Green Street, Urbana, IL 61801}

\date{\today}
 
\begin{abstract}
We investigate different choices for the quadratic momentum dependence
required in nucleon-nucleon potentials to fit phase shifts in
high partial-waves.  In the Argonne~$v_{18}$ potential ${\bf L}^2$ and
$({\bf L\cdot S})^2$ operators are used to represent this dependence.
The $v_{18}$ potential is simple to use in many-body calculations
since it has no quadratic momentum-dependent terms in $S$-waves.
However, ${\bf p}^2$ rather than ${\bf L}^2$ dependence occurs naturally
in meson-exchange models of nuclear forces.  We construct an alternate
version of the Argonne potential, designated Argonne~$v_{18pq}$, in which the
${\bf L}^2$ and $({\bf L\cdot S})^2$ operators are replaced by
${\bf p}^2$ and $Q_{ij}$ operators, respectively.
The quadratic momentum-dependent terms are smaller in the $v_{18pq}$ than
in the $v_{18}$ interaction.  Results for the ground state binding energies of
$^3$H, $^3$He, and $^4$He, obtained with the variational Monte Carlo method,
are presented for both the models with and without three-nucleon interactions.
We find that the nuclear wave functions obtained with the $v_{18pq}$ are
slightly larger than those with $v_{18}$ at interparticle distances $<$ 1 fm.
The two models provide essentially the same binding in the light nuclei,
although the $v_{18pq}$ gains less attraction when a fixed three-nucleon 
potential is added.
\end{abstract}
 

\maketitle

}

\section{Introduction}

A number of choices have to be made in arriving at a form used to 
represent the nucleon-nucleon ($N\!N$) interaction by a model 
potential operator $v_{ij}$.  In general, models with different choices 
will give different results for systems with three and more nucleons. 
When we add many-body forces to the Hamiltonian, they too depend upon the 
choice made for $v_{ij}$.  In principle all realistic combinations of 
$N\!N$ and three-nucleon ($3N$) interaction, $V_{ijk}$, should give 
identical results for the three-nucleon system by construction.  We can 
similarly define combinations of 2-, 3-, 4- and many-nucleon interactions
such that they give the same results for many-nucleon systems.  However, 
such constructions are probably useful only when the contributions of 
$n$-body interactions decrease rapidly with $n$.  In practice the contribution 
of realistic $V_{ijk}$ is much smaller than that of $v_{ij}$, and the 
4- and higher-nucleon terms in the Hamiltonian are often neglected. 

All realistic models of $N\!N$ interaction \cite{SKTS94,WSS95,MSS96}
have a one-pion-exchange (OPE) character at long range, which
is absolutely essential to fit high partial-wave data.   
The choices to be made for the OPE interaction are well known. 
As emphasized by Friar \cite{friar77,coonf86}, they are related by  
unitary transformations which generate the appropriate pion-exchange 
terms in $V_{ijk}$.  For example, a significant part of the difference 
between the Bonn potential \cite{MSS96} and the Nijmegen \cite{SKTS94} 
and Argonne \cite{WSS95} potentials is due to the choice made for the 
OPE interaction \cite{Forest00}. 

In contrast, the origins of the shorter-range parts of the $N\!N$ interaction 
are not well known, and the choices made to model those are more pragmatic.  
When the model is intended for use in many-body problems, the 
accuracy with which it reproduces two-body data and the ease with which it
can be used in calculations are primary considerations \cite{WSS95}.
Phenomenological three-body forces, determined from data \cite{PPWC01},
are used  with these models.  They should reduce the dependence of the 
calculated many-body observables on the choices made to represent $v_{ij}$. 

Realistic models of the shorter-range parts of the $N\!N$ interaction can be 
broadly categorized as phenomenological, meson-exchange, or based on 
effective field theory.  Here we focus on phenomenological models which 
can accurately reproduce the two-nucleon data.  The meson-exchange models 
contain phenomenological effective mesons, and in models 
such as the CD Bonn \cite{MSS96} their masses and couplings depend 
upon the $N\!N$ partial wave.  Most models based on effective field theory 
do not yet fit the scattering data as well. 

Phenomenological models use an operator construct, such as:
\begin{equation}
v_{ij} = \sum_{p=1,n} v_{p}(r_{ij}) O^{p}_{ij} \ .
\end{equation}
They are based on the demonstration~\cite{OM58} that the most 
general form satisfying translational, rotational, parity, and 
time-reversal invariance can be written as:
\begin{eqnarray}
    v_{NN} &=& v_c + v_{\sigma}~{\bf\sigma}_{i}\cdot{\bf\sigma}_{j} 
             + v_t~S_{ij} + v_{ls}~({\bf L\cdot S})_{ij} \nonumber \\
          && + v_{\sigma l}~L_{ij}
 + v_{\sigma{\rm p}}~{\bf\sigma}_{i}\cdot{\bf p}~{\bf\sigma}_{j}\cdot{\bf p} \ ,
\label{eq:nn}
\end{eqnarray}
with the tensor and quadratic spin-orbit operator definitions:
\begin{eqnarray}
    S_{ij} &=& 3~{\bf\sigma}_{i}\cdot{\bf \hat{r}}
                ~{\bf\sigma}_{j}\cdot{\bf \hat{r}} 
               - {\bf\sigma}_{i}\cdot{\bf\sigma}_{j} \\
    L_{ij} &=& \frac{1}{2} \{ {\bf\sigma}_{i}\cdot{\bf L}
                          , {\bf\sigma}_{j}\cdot{\bf L} \} \ .
\end{eqnarray}
Fortunately, on-energy-shell data require only two of the three possible
tensor operators, $S_{ij}$, $L_{ij}$, and 
${\bf\sigma}_{i}\cdot{\bf p}~{\bf\sigma}_{j}\cdot{\bf p}$~.  The last
one has been omitted in all models~\cite{TRS75} since the potential 
associated with it does not have a known theoretical motivation.

The individual $v_x$ in Eq.(\ref{eq:nn}) can be functions of $r$, 
${\bf p}^{2}$, or ${\bf L}^{2}$, and can have a general isospin 
dependence:
\begin{eqnarray}
 v_x &=& v^0_x(r,{\bf p}^{2},{\bf L}^{2})
       + v^{\tau}_x(r,{\bf p}^{2},{\bf L}^{2}){\bf\tau}_{i}\cdot{\bf\tau}_{j} \\
   &+& v^T_x(r,{\bf p}^{2},{\bf L}^{2}) T_{ij} 
    + v^{\tau_z}_x(r,{\bf p}^{2},{\bf L}^{2})(\tau_{zi}+\tau_{zj}) \nonumber \ ,
\end{eqnarray}
where the first two terms are charge-independent (CI), the next has the
charge-dependent (CD) isotensor operator
$T_{ij} = 3 {\bf\tau}_{zi}{\bf\tau}_{zj} - {\bf\tau}_{i}\cdot{\bf\tau}_{j}$,
and the last is a charge-symmetry-breaking (CSB) isovector component.
The nuclear CD and CSB forces are much smaller than the CI terms.

All modern $N\!N$ potentials formulated in an operator structure contain the
eight basic CI operators arising from the first line of Eq.(\ref{eq:nn}):
\begin{equation}
O^{p=1,8}_{ij} = [1, {\bf\sigma}_{i}\cdot{\bf\sigma}_{j}, S_{ij},
{\bf L\cdot S}]\otimes[1,{\bf\tau}_{i}\cdot{\bf\tau}_{j}] \ .
\label{eq:op8}
\end{equation}
These eight operators are sufficient to fit a CI average of $S$- and $P$-waves 
and to obtain good deuteron properties.
Any interaction that is static or has linear terms in momentum can be 
represented as a sum of these eight operators.
Typically six additional operator terms that are quadratic in 
either angular or linear momentum are required to fit higher partial waves.
For example, the Urbana $v_{14}$~\cite{LP81} and Argonne $v_{14}$~\cite{WSA84} 
potentials use the additional operators:
\begin{equation}
O^{p=9,14}_{ij} = [{\bf L}^{2},({\bf\sigma}_{i}\cdot{\bf\sigma}_{j}){\bf L}^{2},
({\bf L\cdot S})^{2}]\otimes[1,{\bf\tau}_{i}\cdot{\bf\tau}_{j}] \ ,
\label{eq:op14}
\end{equation}
where $({\bf L\cdot S})^{2}$ is simply related to $L_{ij}$:
\begin{equation}
    L_{ij} = {\bf L\cdot S} - {\bf L}^{2} + 2 ({\bf L\cdot S})^{2} \ .
\label{eq:lijop}
\end{equation}
The more modern Argonne $v_{18}$ (AV18) potential of Ref.~\cite{WSS95} 
(hereafter referred to as WSS) supplements these fourteen CI operators with
four small CD and CSB terms and a complete electromagnetic interaction.

The Super-Soft-Core (SSC) potential \cite{TS72} also used four 
${\bf L}^{2}$ operators, but with the angular momentum tensor 
operator $Q_{ij}$
\begin{equation}
Q_{ij} = 3 L_{ij} - ({\bf\sigma}_{i}\cdot{\bf\sigma}_{j}){\bf L}^{2} \ ,
\label{eq:qop}
\end{equation}
or, in a later version~\cite{TRS75}, the $L_{ij}$ operator as the specific 
quadratic spin-orbit term.
The $Q_{ij}$ and $({\bf L\cdot S})^{2}$ operators vanish in $S=0$ states, 
while the $L_{ij}$ does not.  
Still other variations of the quadratic spin-orbit operator, 
related via Eqs.(\ref{eq:lijop}) and (\ref{eq:qop}), 
have been used in the past. 

The Urbana, Argonne, and SSC models are all local potentials in each $N\!N$ 
partial wave denoted by quantum numbers $L,~S,~J$.  In single channels having 
$L=J$, the $v_{LSJ}$ is a central function of $r$, while in coupled 
channels having $L=J \pm 1$, the $v_{LSJ}$ is a sum of central, tensor 
and spin-orbit interactions.  
In contrast, the parametrized Paris potential~\cite{Paris80} was constructed 
with ${\bf p}^{2}$ operators instead of ${\bf L}^{2}$, i.e., 
\begin{equation}
   v(r,{\bf p}^{2}) = v^c(r) + {\bf p}^{2}v^{p2}(r)
                               + v^{p2}(r){\bf p}^{2} \ ,
\label{eq:vrp2}
\end{equation}
with
\begin{equation}
    {\bf p}^{2} = -\left[ \frac{1}{r} \frac{d^{2}}{dr^{2}} r 
                      - \frac{{\bf L}^{2}}{r^{2}} \right] \ ,
\label{eq:p2}
\end{equation}
where we have set $\hbar=1$ for brevity. 
Because ${\bf p}^{2}$ does not commute with functions of $r$, a 
momentum-dependent potential like Paris can be more difficult to use in 
many-body calculations, especially configuration-space methods such as quantum
Monte Carlo~\cite{PW01} and variational methods using chain summation 
techniques.  In fact the Urbana model with ${\bf L}^2$ and $({\bf L\cdot S})^2$ 
operators was proposed to avoid difficulties encountered in variational 
calculations of nucleon matter with the Paris potential \cite{LP81b}.
On the other hand, ${\bf p}^{2}$ operators are more likely to arise in
field-theoretic models for the $N\!N$ interaction.

In 1993 the Nijmegen group~\cite{SKRS93} produced a multienergy partial-wave
analysis of elastic $N\!N$ data below $T_{lab} = 350$ MeV (PWA93) that was
able to reproduce over 4300 data points with a $\chi^{2}$/datum $\sim 1$.
This was followed by a series of comparably accurate potential models,
including their own Nijm~I, Nijm~II, and Reid93 potentials~\cite{SKTS94},
AV18~\cite{WSS95}, and CD~Bonn~\cite{MSS96}.
All these models use an operator construct in some fashion, with the
Nijm~II, Reid93, and AV18 choosing operators which give local potentials
in each partial wave, while Nijm~I includes ${\bf p}^2$ terms and CD Bonn
has additional momentum-dependent terms.
However, with the exception of AV18, they are all adjusted partial-wave by
partial-wave.
The isoscalar $v_{14}$ part of the $v_{18}$ potentials contains only
five independent interactions in $S=1$ states and two in $S=0$
states, {\em i.e.}, seven each for the two isospin values $T=0$ and 1.
In the $S=1$ states they correspond to the central,
tensor, spin-orbit, ${\bf L}^2$ or ${\bf p}^2$, and $({\bf L \cdot S})^2$
or $Q_{ij}$, while in the $S=0$ states we have only the central and
${\bf L}^2$ or ${\bf p}^2$ potentials.
In the $T=1,~S=1$ states, for example, the five potentials can be extracted
from the interactions in the $^3P_0,~^3P_1$ and $^3P_2-^3F_2$ partial waves.
The isoscalar interactions in all the $^3F_3$ and higher $T=1,~S=1$ partial
waves are related to those in the $^3P_0,~^3P_1$ and $^3P_2-^3F_2$ in a
$v_{18}$ model.
Similar relations exist in partial waves with $S=0$.
However, these relations are lost when the meson coupling constants or masses
are varied to fit each partial wave, and have different values in each partial
wave, as in the Nijmegen and Bonn models.
These potentials therefore have additional nonlocalities which
are best represented with partial-wave projection operators.

Calculations of trinucleon and alpha binding energies with the five potentials
mentioned above indicate that the three partial-wave local potentials
give similar results, but the other two give more binding~\cite{FPSS93,NKG00}.
CD-Bonn gives 0.4 and 2.0 MeV more binding than AV18 in $^3$H and $^4$He, but 
a large part of it could be due to nonlocal representation of the 
OPE potential used in that model.  As mentioned earlier, this 
extra binding goes into the three-nucleon interactions used with the 
other models.  Nijm~I uses the local OPE potential, but has shorter-range
momentum-dependent terms.  It gives only 0.1 
and 0.7 MeV more binding. However, none of the models is able to reproduce 
the experimental binding
energies without the addition of a three-nucleon potential.

The purpose of this paper is to study the practical consequences of using 
${\bf p}^{2}$ instead of ${\bf L}^{2}$ operators in potential models.
To do this, we construct a variant of the AV18 potential, using the
same spatial functions and refitting the adjustable parameters to make the 
model as phase equivalent to AV18 as possible.
As discussed below, our best reproduction of AV18 phases also involves changing
the quadratic spin-orbit operator from $({\bf L\cdot S})^{2}$ to $Q_{ij}$.
As a result, we designate this new model as Argonne $v_{18pq}$ (AV18pq)
to denote the two changes in operator structure, 
${\bf L}^{2} \rightarrow {\bf p}^{2}$ and 
$({\bf L\cdot S})^{2} \rightarrow Q_{ij}$.

In Sec.~II we describe the construction of the new potential and present a
comparison of phase shifts and radial forms with respect to AV18.
In Sec.~III we present variational Monte Carlo calculations of the $A=3,4$
nuclei for both models, and with different three-nucleon potentials added.
Our conclusions are given in Sec.~IV.

\section{Potential Construction}

The practical need for the quadratic momentum-dependent operators is to
simultaneously fit the phase shifts in $S$- and $D$-waves and in $P$- and 
$F$-waves.  For example, a static potential with components of one- and 
two-pion exchange ranges and a shorter-range core cannot fit the $^1S_0$ 
and $^1D_2$ phase shifts.  The ${\bf L}^2$ dependent term in the $v_{18}$
interaction operator generates a difference between the potentials in 
$^1S_0$ and $^1D_2$ waves, and allows one to fit the phase shifts very 
accurately. 

Quadratic spin-orbit terms are necessary for fitting 
the variety of data in higher $S=1$ partial waves, such as $^3D_{1,2,3}$ and 
$^3F_{2,3,4}$.
The AV18 is written with $({\bf L\cdot S})^{2}$ operators, but it can be
algebraically recast with either the $L_{ij}$ or $Q_{ij}$ operators, with
reshuffling of contributions among the $({\bf L \cdot S})^2$, 
${\bf L\cdot S}$ and ${\bf L}^{2}$ terms.
As discussed below, we find it expedient to choose the $Q_{ij}$ operator, so 
we use AV18 in this representation, which we designate as AV18q for clarity.
This will allow us to study the difference between choosing ${\bf p}^2$ or 
${\bf L}^2$ keeping all other operators the same.  This recast 
potential is the same as AV18, but its ${\bf L}^2$ and ${\bf L}\cdot{\bf S}$ 
terms contain parts of the $({\bf L}\cdot{\bf S})^2$ terms via
Eqs.~(\ref{eq:lijop}) and (\ref{eq:qop}). 

Inserting Eq.(\ref{eq:p2}) in Eq.(\ref{eq:vrp2}) we get
\begin{equation}
    v(r,p^2)=v^c(r)-\{\frac{1}{r} \frac{d^2}{dr^2}r , v^{p2}(r)\} 
            + 2 v^{p2}(r) \frac{L^2}{r^2} \ .
\end{equation}
Therefore, in order to make comparison of terms in AV18q and AV18pq 
meaningful we define:
\begin{equation}
    v^{p2}_{ST}(r) = \frac{1}{2} r^{2} v^{l2}_{ST}(r) \ .
\label{eq:vp}
\end{equation}
The ${\bf L}^2$ term in AV18pq then has the same form as that in AV18q.
Formally any $v_{18pq}$ model is obtained by adding the radial derivatives 
in ${\bf p}^{2}$, as given in Eq.(\ref{eq:p2}), to a $\tilde{v}_{18q}$ 
potential:  
\begin{equation}
   v_{18pq} =\tilde{v}_{18q} 
                     - \{ \frac{1}{r} \frac{d^{2}}{dr^{2}} r , 
		     \frac{1}{2} r^2 v^{l2} \} \ .
\label{eq:schematic}
\end{equation}
However, the parameters of $\tilde{v}_{18q}$ are not equal to those of 
$v_{18q}$ which fits the phase shifts without the term with radial 
derivatives. 

The intermediate- and short-range parts of AV18pq are parameterized as 
in AV18 or equivalently AV18q.  They are given by:
\begin{eqnarray}
    v^{x}_{ST}(r) &=& v^{\pi}_{ST} + I^{x}_{ST} T_{\mu}^{2}(r) \\
       &+& \Big[ P^{x}_{ST} + \mu r\,Q^{x}_{ST}
                + (\mu r)^{2}R^{x}_{ST} \Big] W(r) \nonumber \ .
\end{eqnarray}
Here $x$ can be $c,~t,~ls,~l2$ or $q$ for the central, tensor, spin-orbit,
either ${\bf L}^2$ or ${\bf p}^2$, and $Q_{ij}$ interactions. 
The $v^{\pi}_{ST}$ denotes the OPE interaction which 
contributes to terms with $x=c$ and $t$. 
The average pion mass is $\mu$, $T_{\mu}(r)$ is the tensor
Yukawa function (with a cutoff), and the shape of the intermediate-range
interaction is approximated with $T_{\mu}^2(r)$. A Woods-Saxon 
function $W(r)$ is used to parametrize the short-range part.

The radial wave equation for $u(r)/r$ in single channels with 
the AV18q potential is
\begin{equation}
   \frac{1}{2M_{r}} u^{\prime\prime}(r) 
   = \Big( \frac{1}{2M_{r}} \frac{\ell(\ell+1)}{r^{2}}
        + v_{ch}(r) - E \Big) u(r) \ ,
\label{eq:orig}
\end{equation}
where $v_{ch}(r)$ is the potential in that channel given by the 
$v_{18q}$ operator. 
We solve this equation numerically using the Numerov method, which is
particularly efficient for second-order differential equations without
first derivative terms.
With the addition of the derivative terms of Eq.(\ref{eq:schematic}),
the radial wave equation for AV18pq generalizes to
\begin{eqnarray}
\label{eq:de2}
    && \Big( \frac{1}{2M_{r}} + 2 v^{p2}_{ST}(r) \Big) u^{\prime\prime}(r)
     + 2 [v^{p2}_{ST}(r)]^{\prime} u^{\prime}(r) \\
    &=& \Big( \frac{1}{2M_{r}} \frac{\ell(\ell+1)}{r^{2}}
     + \tilde{v}_{ch}(r) - [v^{p2}_{ST}(r)]^{\prime\prime} 
     - E \Big) u(r) \nonumber \ .
\end{eqnarray}
Although this expression has a first derivative of the wave function in it, 
it can be solved by the Numerov method with a simple substitution:
\begin{equation}
    w(r) = \xi(r) u(r) = [1 + 4M_{r} v^{p2}_{ST}(r)]^{1/2} u(r) \ .
\label{eq:xidef}
\end{equation}
Then Eq.(\ref{eq:de2}) can be rewritten as
\begin{equation}
    D(r) w^{\prime\prime}(r) = \eta(r) w(r) \ ,
\end{equation}
with the definitions
\begin{eqnarray}
\label{eq:dofr}
    D(r)    &=& \frac{1}{2M_{r}} + 2v^{p2}_{ST}(r) \ , \\
\label{eq:etaofr}
    \eta(r) &=& \frac{1}{2M_{r}} \frac{\ell(\ell+1)}{r^{2}} 
                 + \tilde{v}_{ch}(r) 
                 - \frac{([v^{p2}_{ST}(r)]^{\prime})^{2}}{D(r)} - E \ .
\end{eqnarray}

The above method is easily generalized for the coupled channels,
because the new terms of the potential do not couple channels. 
In all other respects, the AV18pq potential is constructed with the same
form as AV18q.
The same set of fundamental constants is used.
The long-range electromagnetic and one-pion-exchange terms are identical,
including their short-range cutoffs.
The strength parameters, $I^{x}_{ST}$, $P^{x}_{ST}$, $Q^{x}_{ST}$, and 
$R^{x}_{ST}$ are refit to make AV18pq as phase equivalent as possible to 
AV18$\equiv$AV18q.  Note that the parameters $P^{x=t}_{ST}$ and 
$Q^{x\neq t}_{ST}$, are determined from the required behavior of 
$v^{x}(r \rightarrow 0)$, as discussed by WSS.

The choice of $Q_{ij}$ as the quadratic spin-orbit operator allows us to
fit the $S=0$ and $S=1$ phases for given $T$ separately. 
The $J\leq 4$ phases are fit for eleven different energies, in the range 
$1-350$ MeV, as well as scattering lengths and deuteron binding.
In the $T=1$ phases, we want the quadratic momentum-dependent parts of the
potential to be charge-independent, so we determine them first by fitting
the phases produced by the CI part of the AV18 potential, including the CI
average of one-pion exchange.
Once these parts are fixed, we adjust the central part of the $S=0$ potential
separately for $pp$ and $np$ phases, including full charge-dependence in
OPE; this sets the CD parts of the interaction.
The $nn$ potential is then slightly altered from the $pp$ potential to 
reproduce the $nn$ scattering length, which is the only constraint on
strong charge-symmetry breaking.
The CSB in $ST=11$ channels is adjusted, as for AV18, such that the final 
potential in operator projection has no spin-dependent CSB, aside from 
electromagnetic terms.
Fitting the $T=0$ channels is straightforward, including reproduction of
the deuteron properties.

The adjusted parameter values for AV18pq are given in Table~\ref{tab:params}, 
in which lines labeled $l2$ give the coefficients of the $v^{l2}_{ST}$ related 
to $v^{p2}_{ST}$ by Eq.(\ref{eq:vp}).  
The rms deviations in degrees of the AV18pq phases from AV18 phases are given
in Table~\ref{tab:pp} for $pp$ scattering, and in Table~\ref{tab:np} for $np$.
These are binned over three different energy ranges, and in total.
To help judge the size of these errors, in the last column we quote the
phase shift errors of PWA93 summed in a similar fashion.

It is apparent from the table that the $S=0$ phases have been reproduced
very well, with rms deviations much smaller than the experimental uncertainty.
The $S=1$ phases are more difficult to reproduce, although the rms deviations
remain generally quite small.
The places where they exceed the PWA93 errors are in the $^{3}$P$_{2}$,
$^{3}$F$_{4}$, and higher $T=1$ phases, and in the $^{3}$S$_{1}$,
$^{3}$D$_{3}$, and $^{3}$G$_{4}$ $T=0$ phases.
The deviations are concentrated at the higher energies.

We also attempted fits using either $({\bf L\cdot S})^{2}$
or $L_{ij}$ as alternate quadratic spin-orbit operators.
Since $({\bf L\cdot S})^{2}$ vanishes in $S=0$ states, the potential in these
channels is the same as in AV18pq.
For the $ST=11$ phases, we were able to get a slightly better reproduction 
of the AV18 model, including $^{3}$P$_{2}$ and $^{3}$F$_{4}$ deviations 
that fall below the PWA93 errors.
However, the $ST=10$ phases were notably worse, with $^{3}$D$_{1}$, 
$^{3}$D$_{2}$, and $^{3}$G$_{4}$ phases having total rms deviations  
$\sim$ 0.3 degrees, which we judge to be unacceptable.
To use $L_{ij}$, we must fit $S=0,1$ channels simultaneously for given $T$,
since this operator does not vanish in $S=0$.
We were able to get an even better reproduction of the AV18 $T=0$ phases
with this operator, but we could not find acceptable $T=1$ solutions.
This does not mean that good ${\bf p}^{2}$ potentials cannot be made with 
the $L_{ij}$ or $({\bf L\cdot S})^{2}$ quadratic spin-orbit operators, but 
only that they may not work so well with the radial functions of the AV18 model.

We now compare the AV18q and AV18pq interactions with each other and with 
the AV8$^\prime$ interaction which has no quadratic momentum-dependent terms.  
It has only the eight leading operators of Eq.(\ref{eq:op8}) plus Coulomb
interaction between protons; it exactly reproduces the AV18 potential in 
$^1S_0$, $^3S_1-^3D_1$ and all the $P$-waves, except for the small CD and 
CSB terms and non-Coulombic electromagnetic terms.
The AV8$^\prime$ is obtained from the AV18 by a simple reprojection of the 
operators~\cite{PPCPW97}, and is not phase equivalent to AV18 and AV18pq 
in $D$- and higher waves.  
Recall that AV18q is obtained by recasting AV18 with the identity:
\begin{equation}
({\bf L}\cdot {\bf S})^2 = \frac{1}{2}\left(1+\frac{1}{3} \sigma_1 \cdot 
  \sigma_j \right) {\bf L}^2 - \frac{1}{2} {\bf L} \cdot {\bf S} + 
  \frac{1}{6} Q_{ij}~,
\end{equation}
and that the only difference between the operator structure of AV18q and 
AV18pq is ${\bf L}^2 \rightarrow {\bf p}^2$. 

The central potentials in $ST=01$ and $10$ channels are compared in 
Fig.~\ref{f:vcsw}.  In these spin-isospin channels the $v^c_{ST}$ are 
those in the $S$ and higher even waves, and they are the same in AV18q and 
AV8$^\prime$ models. 
The $ST=01$ and $10$ central potentials in AV18pq have a slightly smaller 
repulsive core. This is to be expected.  The ${\bf L}^2$
and $Q_{ij}$ operators are zero in $S$-waves.  In the 
AV18q and AV8$^\prime$ models the change in the sign of $S$-wave 
phases at higher energies 
is produced by the repulsive core in the interaction.  In contrast 
the ${\bf p}^2$ part of AV18pq operates in the $S$-waves also.  It becomes more 
repulsive at larger energies, and helps turn the $S$-wave phase shifts negative 
together with the repulsive core.  

\begin{figure}[ht!]
\centering
\includegraphics[width=6.0in]{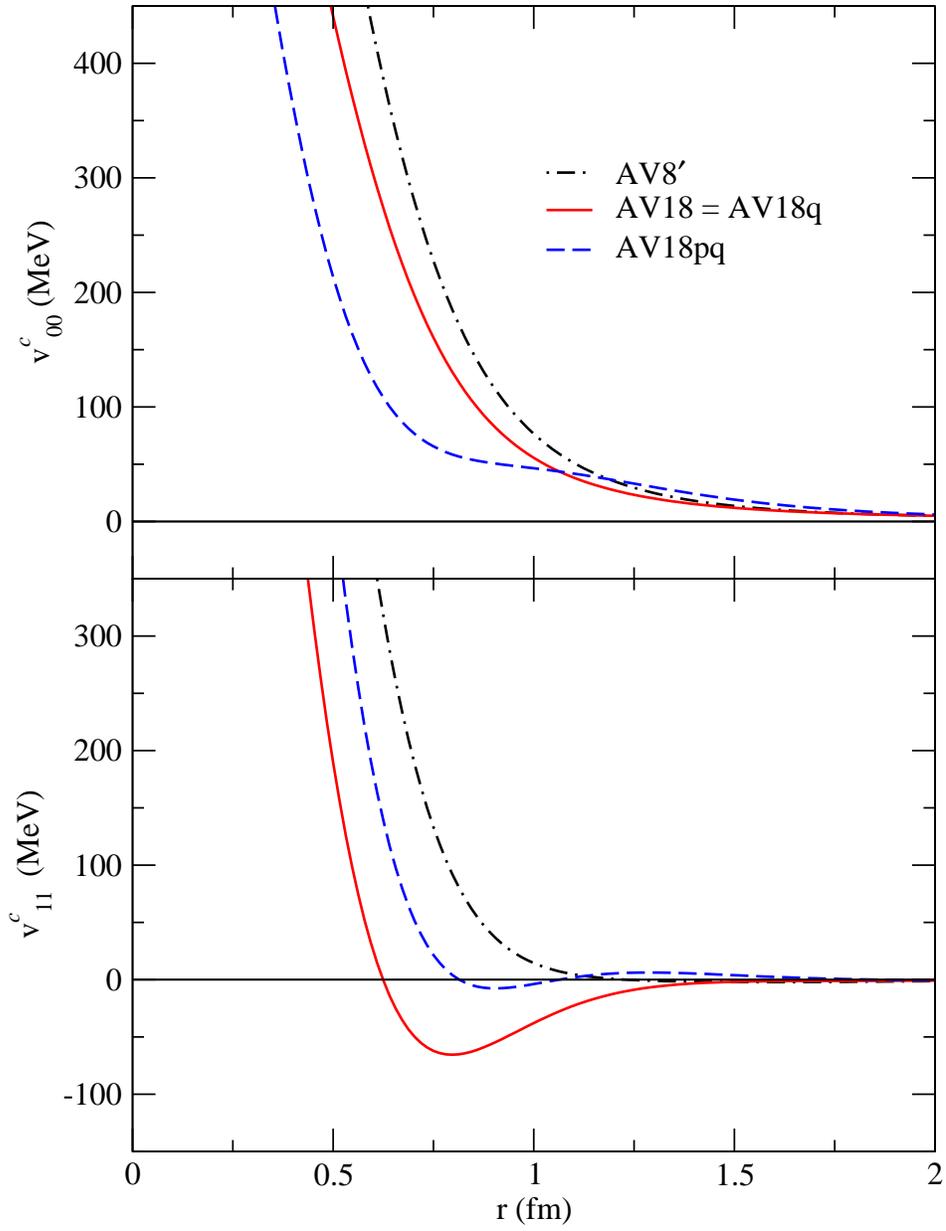}
\caption{(Color online) Central potentials in odd-parity waves.}
\label{f:vcpw}
\end{figure}

The $v^c_{ST}$ in $ST=00$ and $11$ states are compared in Fig.~\ref{f:vcpw}; 
they occur in $P$- and higher odd waves.  In these states the AV8$^\prime$ 
has a larger repulsive core than either AV18 or AV18pq.  The odd 
partial wave $v^c$ in AV8$^\prime$ contain the contribution of 
${\bf L}^2$ and $Q_{ij}$ operators in $P$-waves. 
The repulsive cores in these states are obviously more sensitive to 
the treatment of quadratic momentum dependence as seen by comparing the 
AV18q and AV18pq curves.  They are 
comparable to the centrifugal barrier, $2/(Mr^2)$, in $P$-waves.  For 
example, the centrifugal barrier is $\sim$ 330 MeV at $r=0.5$ fm.  Fortunately 
nuclear energies are not very sensitive to the repulsive cores in $P$- and 
higher waves. 

\begin{figure}[ht!]
\centering
\includegraphics[width=6.0in]{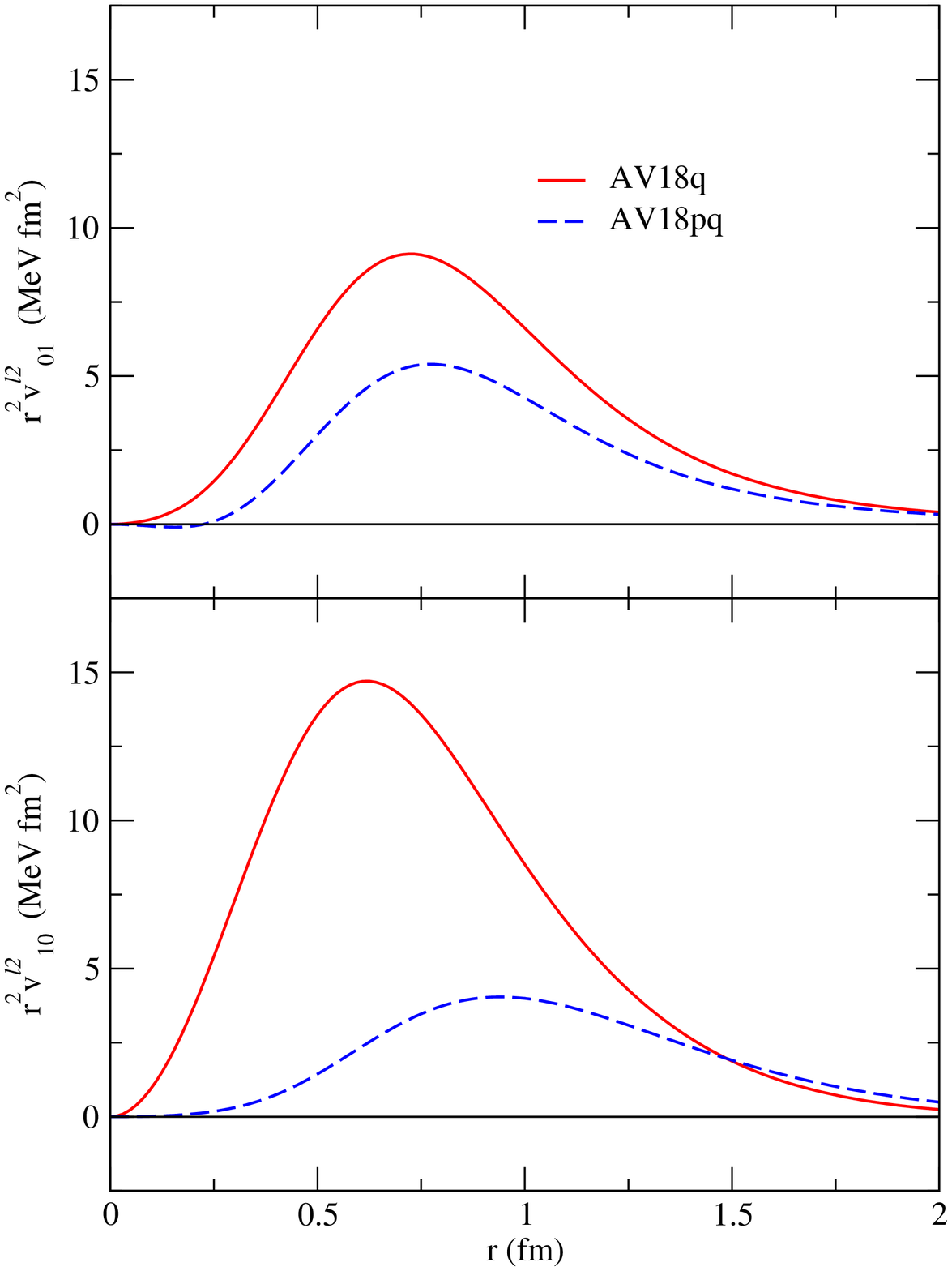}
\caption{(Color online) Quadratic momentum-dependent potentials in even-parity waves.}
\label{f:vl2sw}
\end{figure}

\begin{figure}[ht!]
\centering
\includegraphics[width=6.0in]{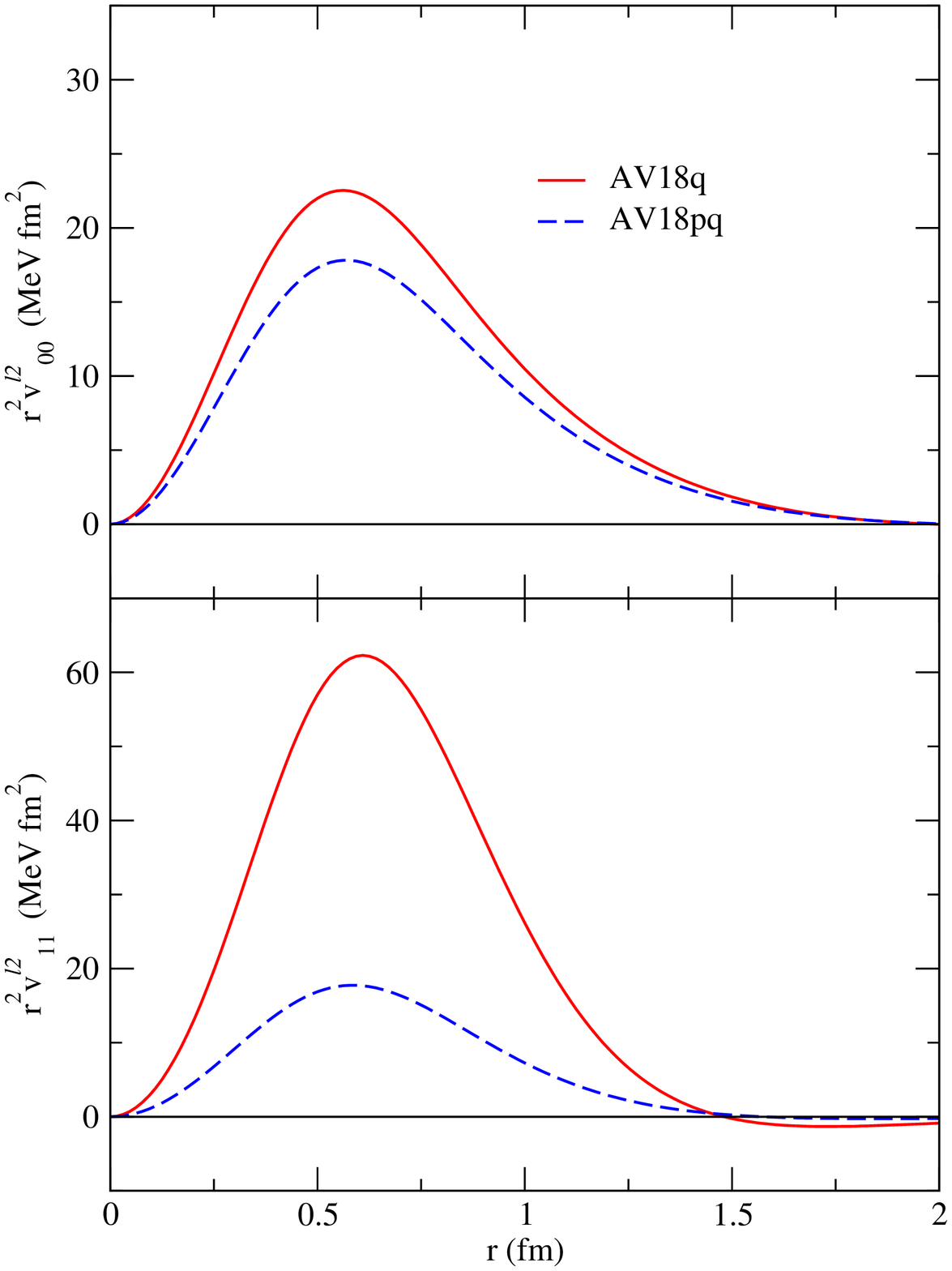}
\caption{(Color online) Quadratic momentum-dependent potentials in odd-parity waves.}
\label{f:vl2pw}
\end{figure}

\begin{figure}[ht!]
\centering
\includegraphics[width=6.0in]{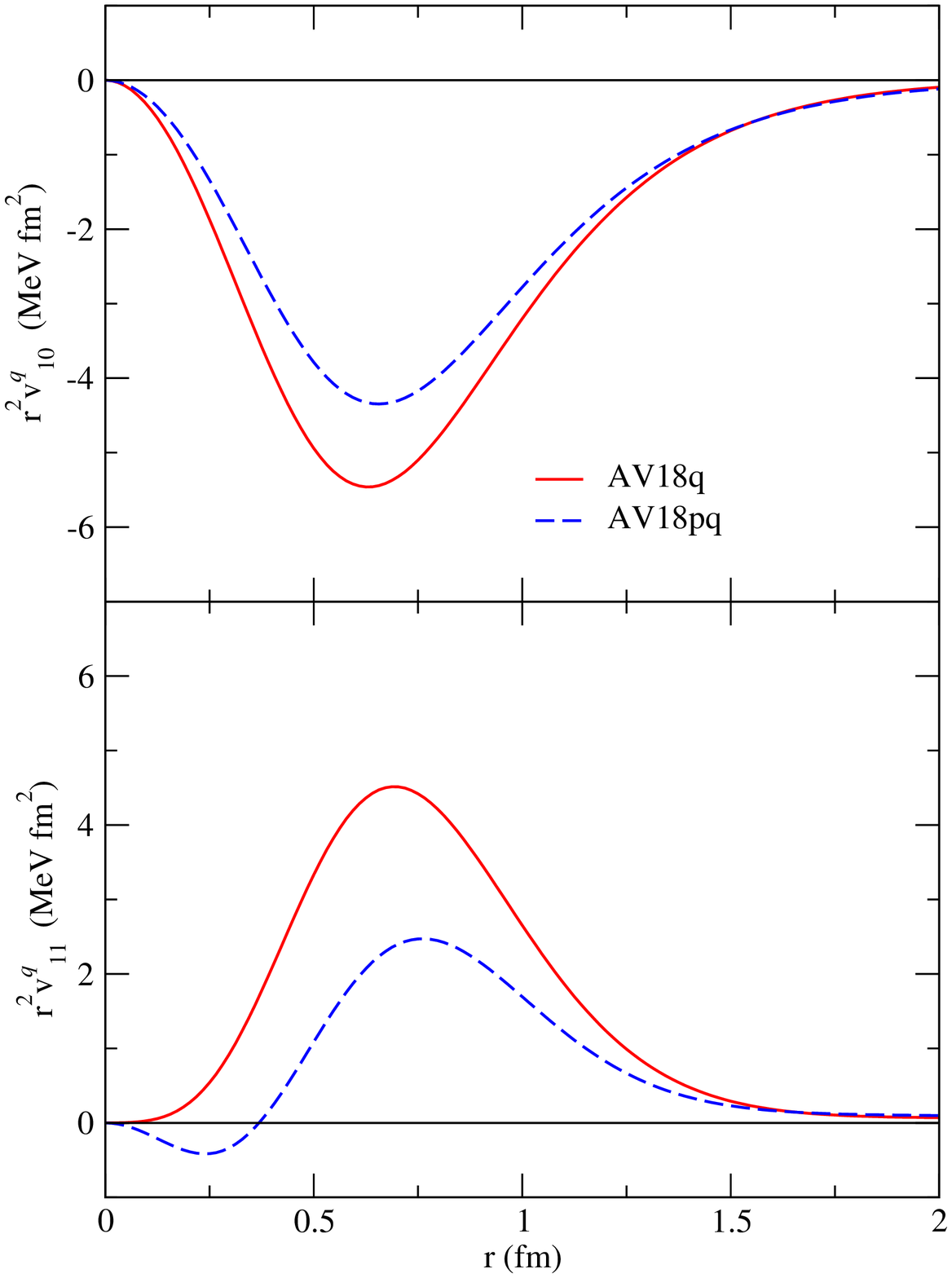}
\caption{(Color online) Quadratic spin-orbit potentials in $S=1$ channels.}
\label{f:vqij}
\end{figure}

The $r^2 v^{l2}_{ST}$ are compared in Figs.~\ref{f:vl2sw} and \ref{f:vl2pw}. 
Recall that the $v^{p2}=r^2 v^{l2}/2$, and that ${\bf L}^2$ also contains 
a $r^2$ factor.  Therefore $r^2 v^{l2}_{ST}(r)$ are the relevant functions. 
In all cases the ${\bf p}^2$ dependent terms in AV18pq are smaller than 
the corresponding ${\bf L}^2$ terms in AV18q.  The $Q_{ij}$ potentials 
are also smaller in AV18pq as shown in Fig.~\ref{f:vqij}. 

\begin{figure}[ht!]
\centering
\includegraphics[width=6.0in]{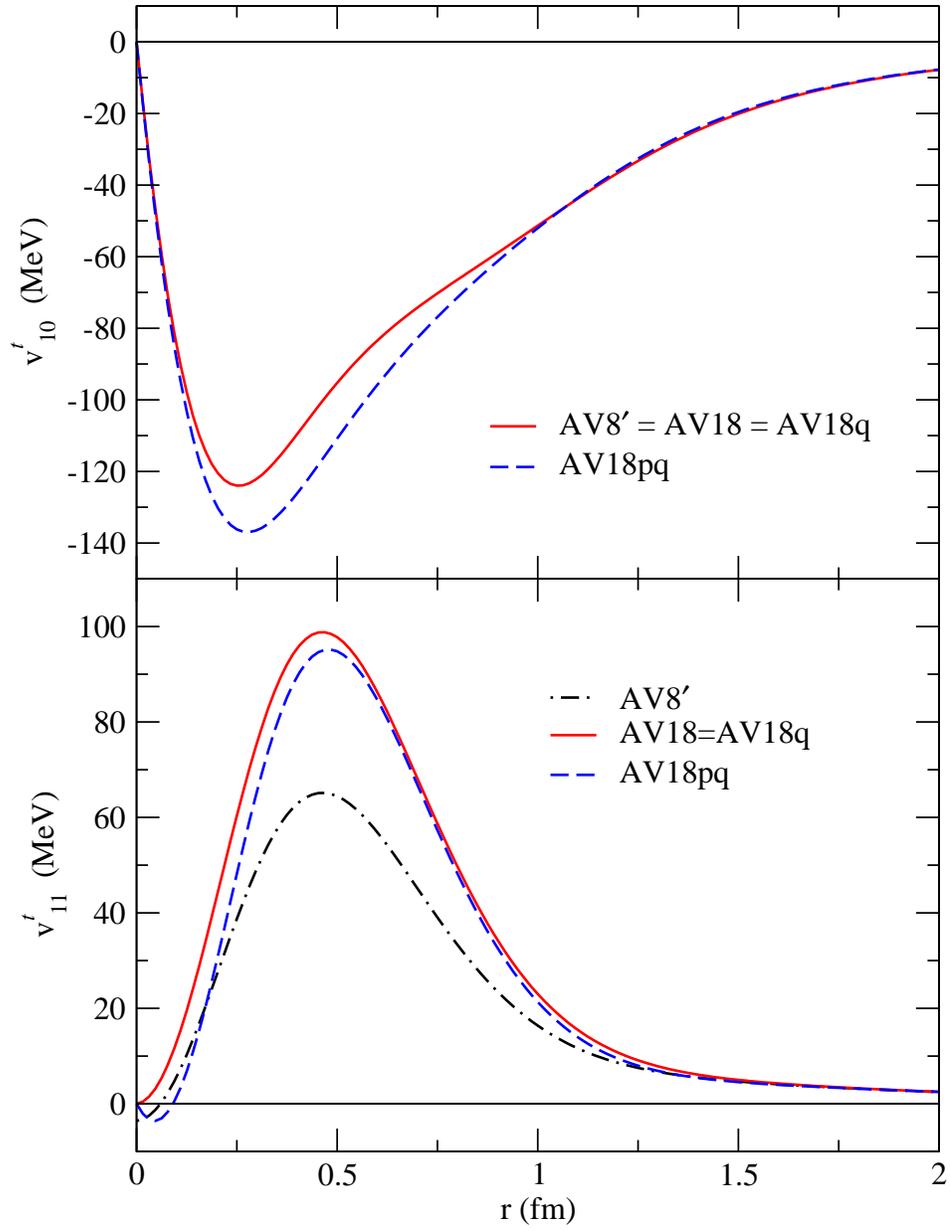}
\caption{(Color online) Tensor potentials in $S=1$ channels.}
\label{f:vt}
\end{figure}

\begin{figure}[ht!]
\centering
\includegraphics[width=6.0in]{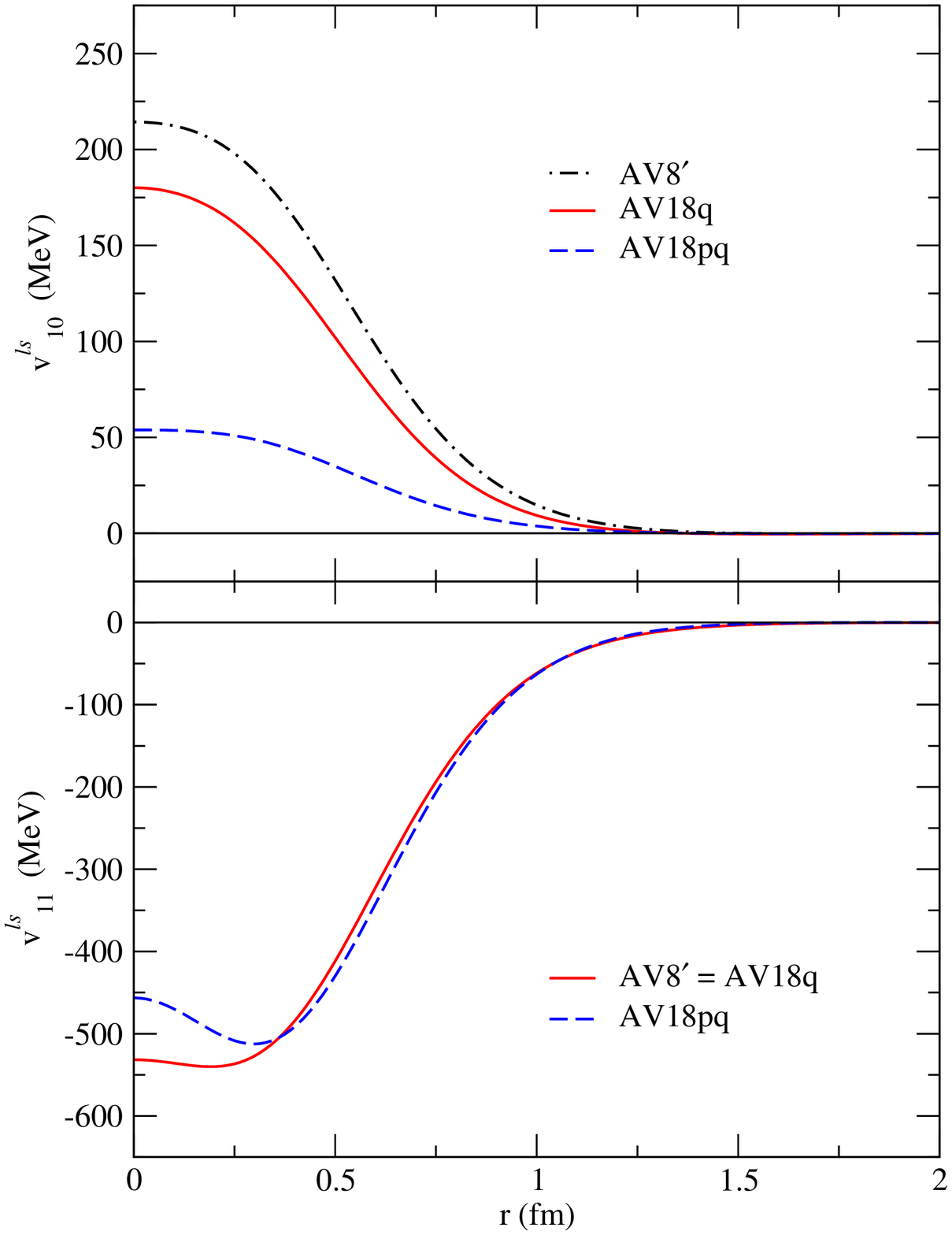}
\caption{(Color online) Spin-orbit potentials in $S=1$ channels.}
\label{f:vls}
\end{figure}

The tensor and spin-orbit potentials are compared in Figs.~\ref{f:vt} and 
\ref{f:vls}.  The tensor potentials are weaker in AV8$^\prime$ but similar in 
AV18q and AV18pq.  The large spin-orbit potential in $ST=11$ states is
similar in all the models.  However, the smaller spin-orbit potential in
$ST=01$ channel is relatively much weaker in AV18pq.

\begin{figure}[ht!]
\centering
\includegraphics[width=6.0in]{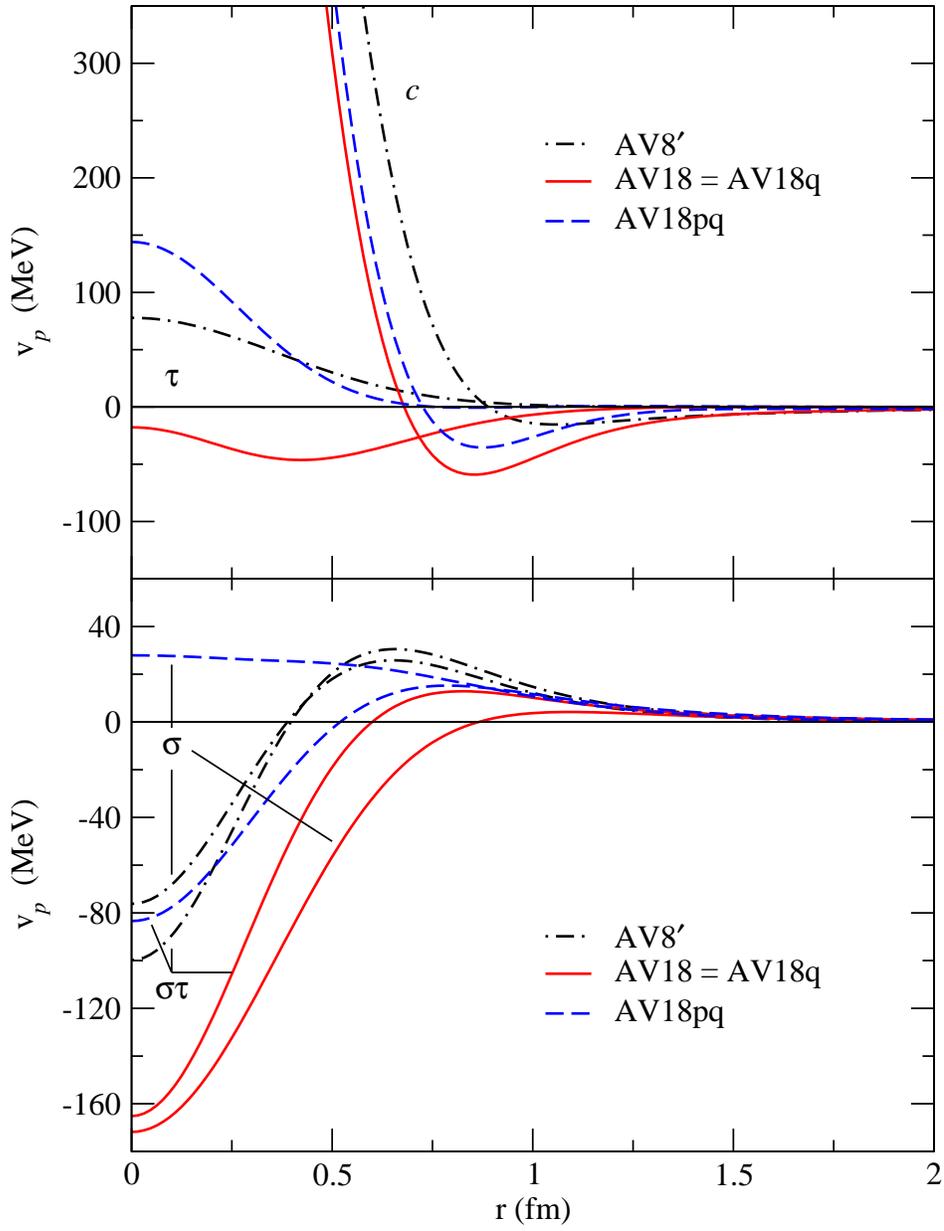}
\caption{(Color online) The $v_{p=1,4}$ projected from the $v^{c}_{ST}$.}
\label{f:v4}
\end{figure}

The potentials in $ST$ states can be cast into an operator format 
using $\sigma_i \cdot \sigma_j$ and $\tau_i \cdot \tau_j$ operators
as detailed in Sec.~IV of WSS.
The $v^c_{ST}$ recast as the sum:
\begin{equation}
v_c(r)+v_{\tau}(r)\tau_i\cdot\tau_j+v_{\sigma}(r)\sigma_i\cdot\sigma_j
+v_{\sigma \tau}(r)\tau_i\cdot\tau_j~\sigma_i\cdot\sigma_j \nonumber
\end{equation}
are shown in Fig.~\ref{f:v4}.  At small $r$, where $v_c$ dominates, the 
smaller $ v_{\tau},~v_{\sigma}$ and $v_{\sigma \tau}$ are poorly 
determined by the data.  The pair current operators depend upon these 
components of the interaction.  For example, in Riska's method~\cite{SPR89} 
the isovector, pseudoscalar and vector currents are obtained from the 
$v_{\sigma \tau}$ and $v_{t \tau}$ potentials associated with the 
$\sigma_i\cdot\sigma_j~\tau_i\cdot\tau_j$ and $S_{ij}~\tau_i\cdot\tau_j$ 
operators.  In particular the $v_{\sigma}$ and $v_{\tau}$ are very different 
in the AV18pq and AV18q models. 

The deuteron properties of AV18q and AV18pq are compared in 
Table~\ref{tab:deut}, along with the expectation values of different 
components of the potential.
The binding energy and other observables are virtually identical.
The total kinetic and potential energies obtained with AV18pq are smaller 
by $\sim$ 1 \%. 
The net contribution of quadratic momentum-dependent terms in AV18pq,
$\langle v^{p2} + v^{q} \rangle = 0.086$ MeV, is noticeably smaller than the
$\langle v^{l2} + v^{q} \rangle = -0.331$ MeV for AV18q.

The deuteron wave functions, $u(r)$ and $w(r)$, are illustrated in 
Fig.~\ref{f:deutwf}; the AV18pq functions peak at slightly smaller values 
of $r$.
Impulse calculations of the deuteron structure function $A(q^{2})$ are
virtually identical, but the $B(q^{2})$ structure function has its minimum
pushed to larger momenta, and likewise the zero crossing of the tensor
polarization $T_{20}(q^{2})$ also increases.
However, consistent two-body current contributions constructed for AV18pq
should be added before the electromagnetic observables are compared in detail
with AV18 predictions or experiment.  

\section{Light Nuclei Calculations}

We have used variational Monte Carlo (VMC) techniques to calculate and compare
the effect of AV8$^\prime$, AV18=AV18q, and AV18pq potentials on the binding of $A=3,4$ 
light nuclei.  AV8$^\prime$ has been used in a benchmark study of four-nucleon binding 
\cite{KNG01} and is the interaction used for propagation in Green's function 
Monte Carlo (GFMC) calculations.
Finally, we study the effects of adding an explicit three-nucleon potential
to the Hamiltonian, either Urbana IX (UIX) \cite{PPCW95} or the TM$^\prime$ 
variant \cite{FHK99} of the Tucson-Melbourne force \cite{CSMBBM79}.
(We use TM$^\prime$ with the cutoff $\Lambda = 4.756 m_{\pi}$ as determined
in Ref.~\cite{NKG00} to be appropriate for use with AV18.)

The VMC calculations are described in Refs.~\cite{W91,APW95}.
We use the trial wave function
\begin{equation}
     |\Psi_V\rangle = \left[1 + \sum_{i<j<k}(U_{ijk}+U^{TNI}_{ijk}) \right]
                      |\Psi_P\rangle \ ,
\label{eq:psiv}
\end{equation}
where the pair wave function, $\Psi_P$, is given by
\begin{equation}
   |\Psi_P\rangle = \left[ {\cal S}\prod_{i<j}(1+U_{ij}) \right]
                   |\Psi_J\rangle \ .
\label{eq:psip}
\end{equation}
The $U_{ij}$, $U_{ijk}$, and $U^{TNI}_{ijk}$ are noncommuting two- and 
three-nucleon correlation operators, and ${\cal S}$ is a symmetrization 
operator.
The antisymmetric Jastrow wave function, $\Psi_J$, depends on the state
under investigation, but for S-shell nuclei takes the simple form:
\begin{equation}
     |\Psi_J\rangle = \left[ \prod_{i<j<k}f^c_{ijk} ({\bf r}_{ik},{\bf r}_{jk})
                             \prod_{i<j}f_c(r_{ij}) \right]
                     |\Phi_A(JMTT_{3})\rangle \ .
\end{equation}
The $f_c(r_{ij})$ and $f^c_{ijk}$ are central (spin-isospin independent) two-
and three-body correlation functions and $\Phi_A$ is an antisymmetric
spin-isospin state, 
\begin{eqnarray}
 &&  |\Phi_{3}(\case{1}{2} \case{1}{2} \case{1}{2} \case{1}{2}) \rangle
        = {\cal A} |\uparrow p \downarrow p \uparrow n \rangle \ , \\
 &&  |\Phi_{4}(0 0 0 0) \rangle
        = {\cal A} |\uparrow p \downarrow p \uparrow n \downarrow n \rangle \ ,
\end{eqnarray}
with ${\cal A}$ the antisymmetrization operator.

The two-body correlation operator $U_{ij}$ is a sum of spin, isospin, and
tensor terms:
\begin{equation}
   U_{ij} = \sum_{p=2,6} \left[ \prod_{k\not=i,j}f^p_{ijk}({\bf r}_{ik}
              ,{\bf r}_{jk}) \right] u_p(r_{ij}) O^p_{ij} \ ,
\end{equation}
where the $O^p_{ij}$ are the first six operators of Eq.(\ref{eq:op8}).
The $f^p_{ijk}$ is another central three-body correlation function.
In previous calculations we also introduced a spin-orbit correlation in
the sum of Eq.(\ref{eq:psiv}), but it is expensive to calculate and improves
the binding energy only a little; we neglect such correlations here.

The central $f_c(r)$ and noncentral $u_p(r)$ pair correlation
functions reflect the influence of the two-body potential at short distances,
while satisfying asymptotic boundary conditions of cluster separability.
Reasonable functions are generated by minimizing the two-body cluster
energy of a somewhat modified $N\!N$ interaction;
this results in a set of coupled, Schr\"{o}dinger-like, differential
equations corresponding to linear combinations of the operators
in $v_{ij}$, with a number of embedded variational parameters \cite{W91}.

For the AV18pq potential, these differential equations must be generalized
to take into account the influence of the $v^{p2}$ terms.
However, by studying the changes in the Schr\"{o}dinger equation for the
phase shifts, Eqs.(\ref{eq:orig}-\ref{eq:etaofr}) above, we find three
simple substitutions that generalize the correlation generators,
Eqs.(2.2-2.5) of Ref.~\cite{W91}, to produce the appropriate correlation 
functions 
$f_{ST}$:
\begin{eqnarray}
   \frac{1}{2M_{r}} &\rightarrow& D(r) \ , \nonumber \\
   v_{ch}(r) &\rightarrow& \tilde{v}_{ch}(r) 
             - \frac{([v^{p2}_{ST}(r)]^{\prime})^{2}}{D(r)} \ ,
                                                   \nonumber \\
   f_{ST}(r) &\rightarrow& f_{ST}(r)/\xi(r) \ . \nonumber
\end{eqnarray}
A reprojection of the $f_{ST}(r)$ into operator terms then yields the 
$f_c(r)$ and $u_p(r)$.

The $f^c_{ijk}$, $f^p_{ijk}$, and $U_{ijk}$ are three-nucleon correlations
also induced by $v_{ij}$; details may be found in Ref.~\cite{APW95}.
The $U^{TNI}_{ijk}$ are three-body correlations induced by the three-nucleon
interaction, which have the form suggested by perturbation theory:
\begin{equation}
     U^{TNI}_{ijk} = \sum_x \epsilon_x V_{ijk}(\tilde{ r}_{ij},
                     \tilde{r}_{jk}, \tilde{ r}_{ki}) \ ,
\label{eq:bestuijk}
\end{equation}
with $\tilde{r}=yr$, $y$ a scaling parameter, and $\epsilon_x$ a (small
negative) strength parameter.

Results of our VMC calculations are given in Table~\ref{tab:vmc}.
We also show GFMC \cite{PPWC01} and Faddeev-Yakubovsky (FY) \cite{NKG00} 
results for comparison, where available.
The VMC results are typically 1.5--3\% above the more exact GFMC and FY
results, which agree with each other to better than 1\%.
The primary goal here is to judge the relative binding of the different
Hamiltonians, and we think the VMC calculations are adequate for this purpose.

In all cases we see that the AV8$^\prime$, with no quadratic momentum 
dependence, gives the most binding.
Both AV18 and AV18pq give less binding and the difference between their
results is smaller than that from AV8$^\prime$.  From the point of view 
of fitting NN data, adding quadratic momentum-dependent terms to $v_{ij}$ is 
the most important step, whether they be ${\bf p}^2$ or ${\bf L}^2$.
In A=3 and 4 nuclei these terms reduce the binding by $\sim$ 2.5 and 4 \%
respectively. 
In the absence of three-nucleon interactions the reductions are nearly the same 
for AV18 and AV18pq.
(In examining the $A=3$ results, it is important to remember that aside from
Coulomb interaction, AV8$^\prime$ is a charge-independent potential, so 
averages of the $^3$H and $^3$He energies show the differences more clearly.)

When a three-nucleon interaction is added to the Hamiltonian, as required
to fit the experimental binding of light nuclei, the effect of adding
quadratic momentum dependence grows to $\sim$ 3 and 5 \% for A=3 and 4
if AV18 is used, while a slightly larger effect of $\sim$ 4 and 6 \% is
obtained with AV18pq.
In practice the strengths of the various terms in the three-nucleon interaction
are not known {\em a priori}.
The short-range terms of this interaction depend upon the representation of
$v_{ij}$, and can be different for AV18 and AV18pq.
They should be chosen separately such that both models give the same,
experimental trinucleon energy.

We note that this interplay between two- and three-nucleon interactions
was already apparent in the Bochum FY calculations \cite{NKG00} for
the Nijmegen potentials \cite{SKTS94}.
These results are also shown at the bottom of Table~\ref{tab:vmc}.
The nonlocal Nijm~I potential gives slightly more binding than the local
Nijm~II, but to obtain the experimental binding by adding a $3N$ force,
a stronger version (larger $\Lambda$ $\rightarrow$ weaker cutoff) is required
to accompany the nonlocal $N\!N$ potential, in this case chosen to reproduce 
the binding of $^3$He.

\section{Conclusions}

We have constructed an AV18pq two-nucleon interaction that is essentially 
phase equivalent to AV18.  The difference between the two interaction 
operators is the choice made for quadratic momentum-dependent terms.  The 
AV18pq uses the ${\bf p}^2$ operator instead of the ${\bf L}^2$ used in AV18.
The results obtained with these two interaction models for the 
$A=2,3$, and 4 bound states are compared with those of the AV8$^\prime$ model 
which has no quadratic momentum dependence.

\begin{figure}[ht!]
\centering
\includegraphics[width=6.0in]{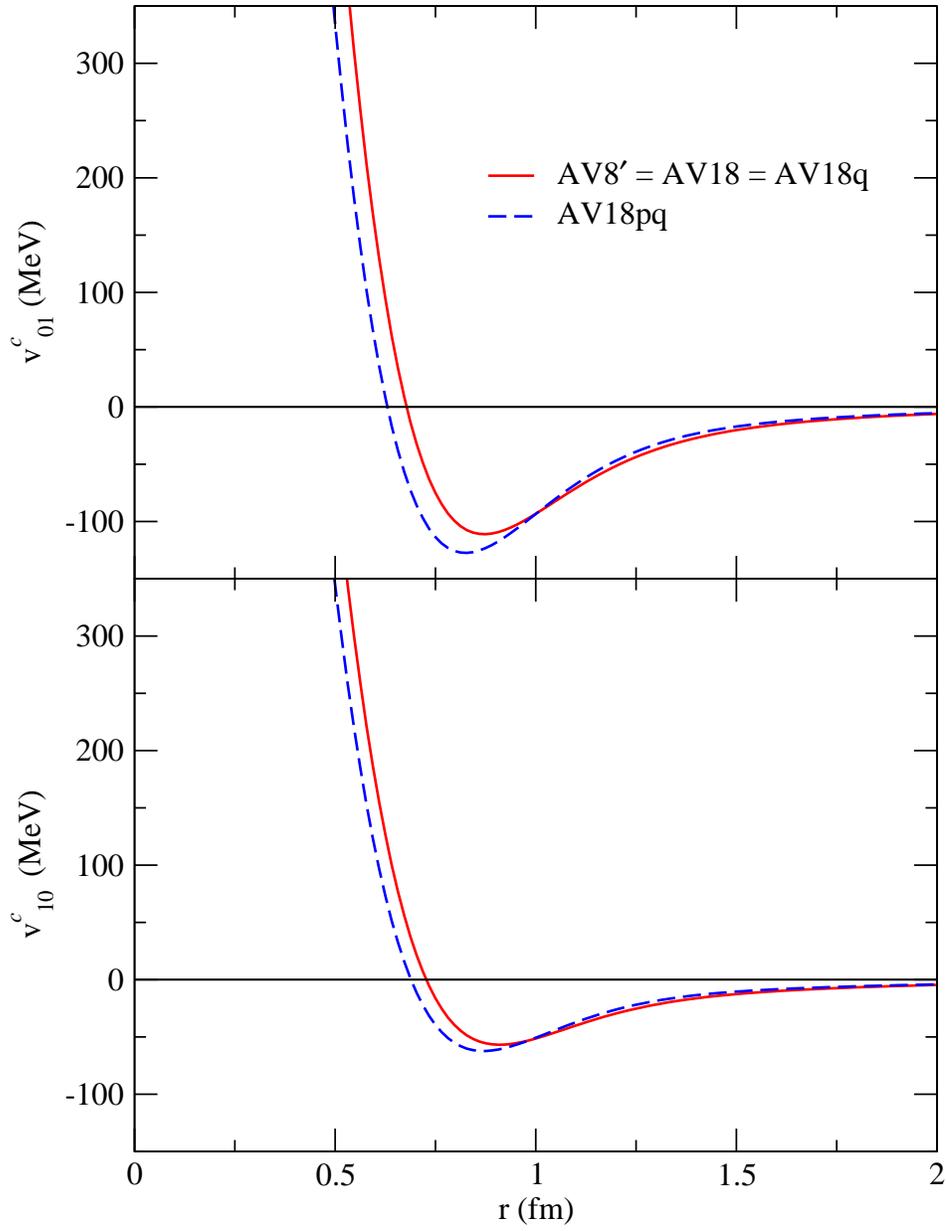}
\caption{(Color online) Central potentials in even-parity waves.}
\label{f:vcsw}
\end{figure}

\begin{figure}[ht!]
\centering
\includegraphics[width=6.0in]{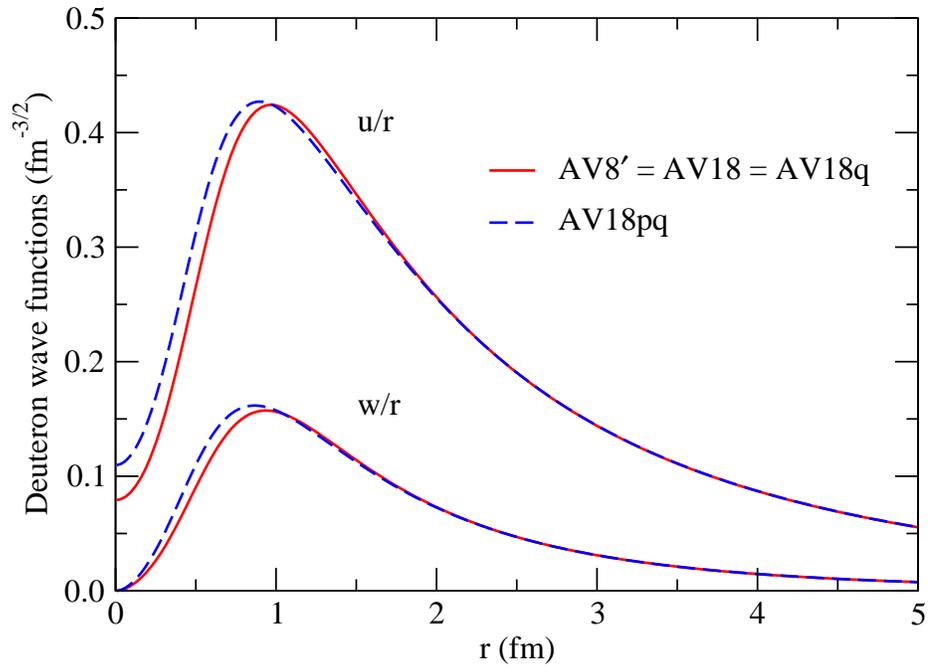}
\caption{(Color online) Deuteron $S$- and $D$-wave functions.}
\label{f:deutwf}
\end{figure}

\begin{figure}[ht!]
\centering
\includegraphics[width=6.0in]{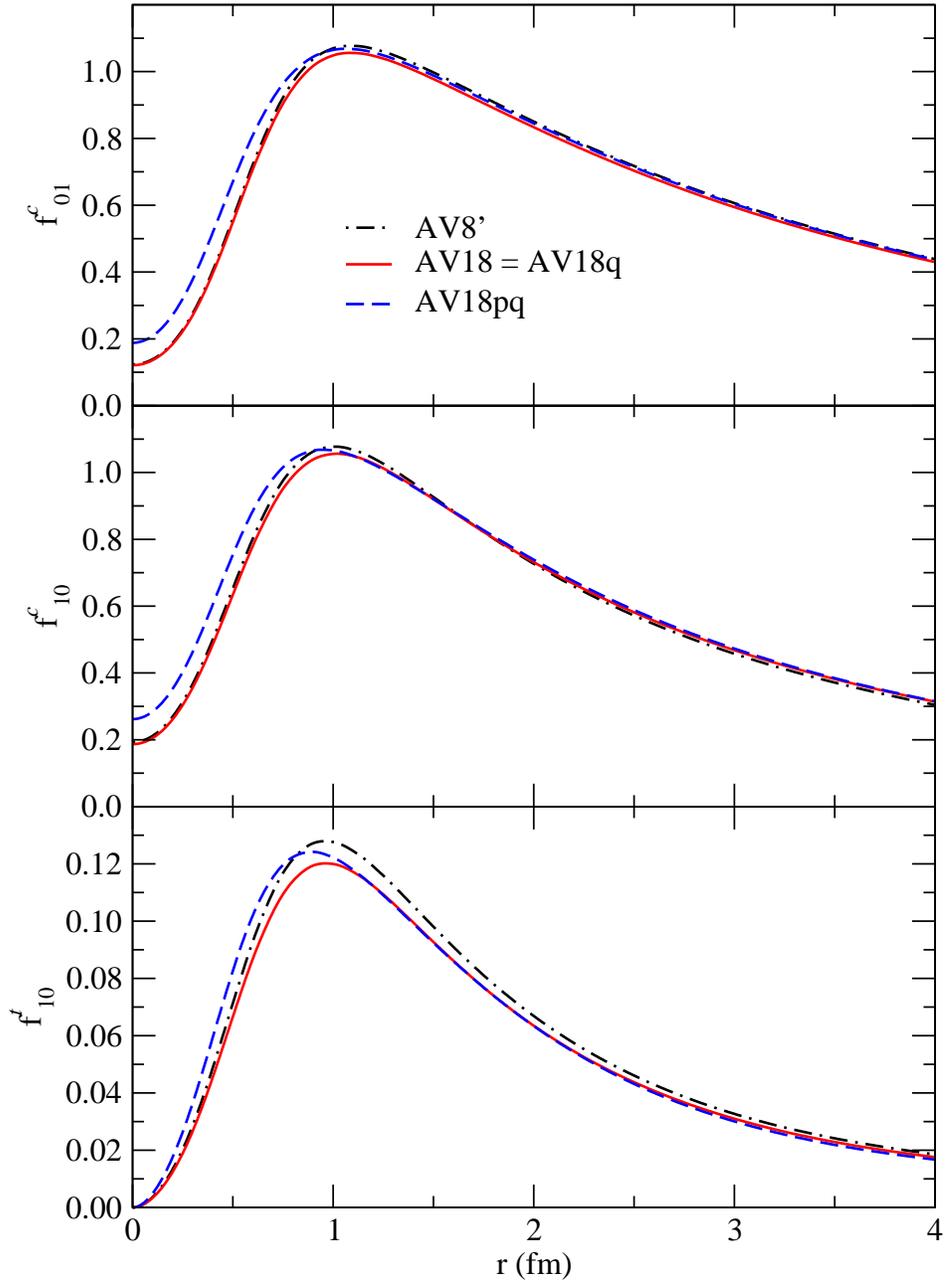}
\caption{(Color online) Correlation functions for $S$-wave channels in $^4$He.}
\label{f:corr}
\end{figure}

The differences in the predicted deuteron wave functions are shown in 
Fig.~\ref{f:deutwf}.  
Those in the $^4$He wave function can be judged from differences in the 
dominant central correlations in $ST=01$, and central and tensor 
correlations in $ST=10$ states shown in Fig.~\ref{f:corr}.  
In all these the magnitude of the wave function predicted by AV18pq 
is larger than that of AV18 or AV8$^\prime$ 
at small interparticle distance $r$.  This is due to the smaller repulsive 
cores in the $S$-waves of AV18pq shown in Fig.~\ref{f:vcsw}. 

In principle the deuteron form factors predicted by AV18pq will be 
different from those of AV18.  We plan to calculate the current 
operators for AV18pq following the methods used for AV18 \cite{WSS95}, and 
compare the predicted form factors with available data. 
Such a comparison could indicate if one of the two choices is more realistic. 

The quadratic momentum-dependent terms give a small positive contribution to
the ground-state energies of nuclei as can be seen from Table~\ref{tab:vmc}.
This contribution can depend upon the choice of interaction operators.
In the absence of a $3N$ interaction, the quadratic momentum-dependent terms
in AV18 and AV18pq give almost identical corrections, within 0.5\%.
However, a larger difference between AV18 and AV18pq develops when a fixed
realistic $3N$ interaction is added to the nuclear Hamiltonian to obtain 
binding energies closer to the experimental values.  
In this case, the non-local AV18pq becomes less attractive than the local
AV18, by $\sim$ 1\%, just as the non-local Nijm~I became relatively less 
attractive compared to the local Nijm~II potential.
One can presumably choose the parameters of the $V_{ijk}$ to reproduce binding 
energies of nuclei with either AV18 or AV18pq. 

All one-boson exchange models of nuclear forces give interactions with a 
${\bf p}^2$ dependence.  Electromagnetic and color forces also have 
a ${\bf p}^2$ dependence.  On the other hand, nucleons are composite 
objects with internal excitations, and their interactions 
can have either ${\bf L}^2$ or ${\bf p}^2$ terms or both.
The main motivation behind the choice of ${\bf L}^2$ 
was to simplify many-body calculations \cite{LP81b}.  
Green's function Monte Carlo methods can be used to study nuclei having 
$A \leq 12$ with AV8$^\prime$ type interactions that contain the dominant 
features of nuclear forces, but not the quadratic momentum dependence. 

In GFMC studies with the AV18 interaction, the quadratic momentum-dependent 
terms are treated as a first-order perturbation \cite{PPWC01}.  
The error $\delta E$ in this calculation is given by: 
\begin{equation}
\delta E = \langle \Psi_{18}|H_{18}| \Psi_{18} \rangle - 
           \langle \Psi_{8^\prime}|H_{18}| \Psi_{8^\prime} \rangle ~, 
\end{equation}
where $| \Psi_{18} \rangle$ and $| \Psi_{8^\prime} \rangle $ are the eigenstates
obtained with the Hamiltonian $H_{18}$ having the AV18 interaction and with 
$H_{8^\prime}$ having the AV8$^\prime$ respectively.  The Hamiltonians can 
include any static 3N interaction.  
The values obtained for $\delta E$ in $^4$He, using 
optimum variational wave functions to approximate the 
$| \Psi_{18} \rangle$ and $| \Psi_{8^\prime} \rangle $ are respectively 
0.05, 0.16 and 0.03 MeV for AV18 alone and with UIX and TM$^\prime$ models of 
3N interaction. 
They are less than 1\% of the total energy. 

A similar study using AV18pq instead of AV18 gives much larger estimates 
of 1.6, 1.1 and 1.8 MeV for the $\delta E$ in the above three cases. 
These larger values are due to differences in the wave functions of 
AV8$^\prime$ and AV18pq models at small $r$ (see Figs.~\ref{f:deutwf} and 
\ref{f:corr}).  A first-order 
treatment of the difference between AV18pq and AV8$^\prime$ gives much too 
large an effect.  It will be necessary to develop new techniques to perform 
accurate GFMC many-body calculations with the AV18pq interaction.  

For those who wish to use the AV18pq potential model in their own applications,
a {\sc fortran} subroutine is available online \cite{fortran}.

\newpage

We thank S. C. Pieper for useful discussions.
This work is supported by the U. S. Department of Energy, 
Nuclear Physics Division, under contract No. W-31-109-ENG-38,
and the U. S. National Science Foundation via Grant No. PHY~00-98353.

%\vfill
\newpage

%\widetext

\begin{table}
\caption{Short-range potential parameters in MeV. The asterisk denotes that
the value was computed by Eq.~(24) of WSS and not fit. The
three shape parameters are: $c=2.1$ fm$^{-2}$, $r_{0}=0.5$ fm, and $a=0.2$ fm.}
\begin{tabular}{ccdddd}
   Channel       & Type &    I       &     P     &      Q         &     R     \\
\colrule
 $S=0, T=1 (pp)$ & $c$  & -10.518030 & 2836.0715 &  1582.7028\ast &  651.1945 \\
 $S=0, T=1 (np)$ & $c$  & -10.812190 & 2816.4190 &  1578.2721\ast & 1002.5300 \\
 $S=0, T=1 (nn)$ & $c$  & -10.518030 & 2832.4903 &  1580.7610\ast &  651.1945 \\
 $S=0, T=1$      & $l2$ &   0.134747 &   -9.4691 &    -5.1342\ast &    0  \\
                 &      &            &           &                &       \\
 $S=0, T=0$      & $c$  &  -4.739629 & 1121.2225 &   466.8222\ast & 2764.3395 \\
                 & $l2$ &  -0.227084 &  166.5629 &    90.3117\ast &    0  \\
                 &      &            &           &                &       \\
 $S=1, T=1 (pp)$ & $c$  &  -9.882847 & 2589.7742 &  1389.2081\ast & 2952.3910 \\
 $S=1, T=1 (np)$ & $c$  &  -9.882847 & 2587.9836 &  1386.1622\ast & 2952.3910 \\
 $S=1, T=1 (nn)$ & $c$  &  -9.882847 & 2586.1930 &  1387.2664\ast & 2952.3910 \\
 $S=1, T=1$      & $l2$ &  -0.008159 &  132.7694 &    71.9886\ast & -169.8510 \\
                 & $t$  &   1.420069 &    0      &  -453.5357     & -837.3820 \\
                 & $ls$ &  -1.749197 & -493.8470 &  -267.7677\ast & 1533.0637 \\
                 & $q$  &   0.135181 &  -17.7975 &    -9.6499\ast &  -46.2542 \\
                 &      &            &           &                &      \\
 $S=1, T=0$      & $c$  &  -8.351808 & 2325.5929 &  1307.9923\ast &  957.8091 \\
                 & $l2$ &  -0.023577 &    1.8164 &     0.9849\ast &  127.2921 \\
                 & $t$  &   1.327862 &    0      & -1170.8528     &  580.5596 \\
                 & $ls$ &   0.060223 &   58.3208 &    31.6220\ast & -126.0235 \\
                 & $q$  &   0.000589 &  -25.1123 &   -13.6161\ast &   -4.6897
\end{tabular}
\label{tab:params}
\end{table}

\begin{table}
\caption{RMS deviations of $pp$ phases in degrees for AV18pq compared to AV18.}
\begin{tabular}{cccccc}
   Channel          & 1-25  & 50-150 & 200-350 & Total & PWA93 \\
                    &  MeV  &   MeV  &    MeV  &       &       \\
\colrule
 $^{1}S_{0}$        & 0.021 &  0.031 &   0.245 & 0.149 & 0.181 \\
 $^{1}D_{2}$        & 0.001 &  0.011 &   0.061 & 0.037 & 0.055 \\
 $^{1}G_{4}$        & 0     &  0.001 &   0.020 & 0.012 & 0.029 \\
 $^{3}P_{0}$        & 0.024 &  0.024 &   0.092 & 0.059 & 0.224 \\
 $^{3}P_{1}$        & 0.002 &  0.035 &   0.074 & 0.048 & 0.109 \\
 $^{3}P_{2}$        & 0.008 &  0.090 &   0.160 & 0.107 & 0.062 \\
 $\varepsilon_{2}$  & 0.002 &  0.011 &   0.044 & 0.027 & 0.048 \\
 $^{3}F_{2}$        & 0.001 &  0.013 &   0.055 & 0.034 & 0.062 \\
 $^{3}F_{3}$        & 0.002 &  0.042 &   0.034 & 0.030 & 0.054 \\
 $^{3}F_{4}$        & 0.002 &  0.054 &   0.077 & 0.054 & 0.043 \\
 $\varepsilon_{4}$  & 0     &  0.005 &   0.009 & 0.006 & 0.001 \\
 $^{3}H_{4}$        & 0     &  0.018 &   0.047 & 0.030 & 0     \\
\end{tabular}
\label{tab:pp}
\end{table}

\begin{table}
\caption{RMS deviations of $np$ phases in degrees for AV18pq compared to AV18.}
\begin{tabular}{cccccc}
   Channel          & 1-25  & 50-150 & 200-350 & Total & PWA93 \\
                    &  MeV  &   MeV  &    MeV  &       &       \\
\colrule
 $^{1}S_{0}$        & 0.018 &  0.087 &   0.078 & 0.067 & 0.344 \\
 $^{1}P_{1}$        & 0.005 &  0.029 &   0.064 & 0.041 & 0.160 \\
 $^{1}F_{3}$        & 0     &  0.003 &   0.022 & 0.013 & 0.047 \\
 $^{3}S_{1}$        & 0.095 &  0.113 &   0.387 & 0.248 & 0.154 \\
 $\varepsilon_{1}$  & 0.011 &  0.061 &   0.048 & 0.043 & 0.115 \\
 $^{3}D_{1}$        & 0.002 &  0.086 &   0.200 & 0.129 & 0.127 \\
 $^{3}D_{2}$        & 0     &  0.037 &   0.139 & 0.086 & 0.090 \\
 $^{3}D_{3}$        & 0.001 &  0.017 &   0.230 & 0.139 & 0.069 \\
 $\varepsilon_{3}$  & 0.001 &  0.005 &   0.036 & 0.022 & 0.035 \\
 $^{3}G_{3}$        & 0     &  0.001 &   0.004 & 0.002 & 0.002 \\
 $^{3}G_{4}$        & 0     &  0.027 &   0.253 & 0.153 & 0     \\
\end{tabular}
\label{tab:np}
\end{table}

\begin{table}
\caption{Comparison of static deuteron properties and expectation values 
for different terms in the potentials.}
\begin{tabular}{cdddc}
                          & $Expt$      &   $AV18q$  &   $AV18pq$ &  \\
\colrule
   $E_{d}$                & 2.224575(9) &   2.224575 &   2.224573 &  MeV  \\
 $\langle T\rangle$       &             &  19.810    &  19.525    &  MeV  \\
 $\langle V\rangle$       &             & -22.035    & -21.750    &  MeV  \\
 $\langle v^{c}\rangle$   &             &  -4.446    &  -4.079    &  MeV  \\
 $\langle v^{t}\rangle$   &             & -16.586    & -17.462    &  MeV  \\
 $\langle v^{ls}\rangle$  &             &  -0.690    &  -0.313    &  MeV  \\
 $\langle v^{l2}\rangle$  &             &   1.087    &            &  MeV  \\
 $\langle v^{p2}\rangle$  &             &            &   1.440    &  MeV  \\
 $\langle v^{q}\rangle$   &             &  -1.418    &  -1.354    &  MeV  \\
 $\langle v^{EM}\rangle$  &             &   0.018    &   0.018    &  MeV  \\
   $A_{S}$                & 0.8781(44)  &   0.8850   &   0.8863   &  fm$^{1/2}$ \\
   $\eta$                 & 0.0256(4)   &   0.0250   &   0.0250   &       \\
   $r_{d}$                & 1.953(3)    &   1.967    &   1.969    &  fm   \\
   $\mu_{d}$              & 0.857406(1) &   0.847    &   0.847    & $\mu_{0}$ \\
   $Q_{d}$                & 0.2859(3)   &   0.270    &   0.270    &  fm$^{2}$ \\
   $P_{d}$                &             &   5.76     &   5.78     &  \%
\end{tabular}
\label{tab:deut}
\end{table}

\begin{table}
\caption{Binding energies of light nuclei (in MeV) for different Hamiltonians
as computed by the variational Monte Carlo (VMC) method, with comparisons
to relevant Green's function Monte Carlo (GFMC) results of Ref.~\cite{PPWC01}
and Faddeev-Yakubovsky (FY) results of Ref.~\cite{NKG00}.  We also show FY
results for the Nijmegen potential at the bottom.  Different versions of
the Tucson-Melbourne (TM) force are characterized by the cutoff parameter
$\Lambda$ shown in parentheses.}
\begin{tabular}{llccc}
 Hamiltonian       & Method      & $^{3}$H    & $^{3}$He   & $^{4}$He    \\
\colrule
 AV18              & VMC         & $-$7.50(1) & $-$6.77(1) & $-$23.70(2) \\
                   & GFMC        & $-$7.61(1) & $-$6.87(1) & $-$24.07(4) \\
                   & FY          & $-$7.623~~ & $-$6.924~~ & $-$24.28~~~ \\
 AV18pq            & VMC         & $-$7.50(1) & $-$6.77(1) & $-$23.79(2) \\
 AV8$^\prime$      & VMC         & $-$7.65(1) & $-$7.01(1) & $-$24.69(2) \\
                   & GFMC        & $-$7.76(1) & $-$7.12(1) & $-$25.14(2) \\
                                 &            &            &             \\
 AV18/UIX          & VMC         & $-$8.29(1) & $-$7.53(1) & $-$27.58(2) \\
                   & GFMC        & $-$8.46(1) & $-$7.71(1) & $-$28.33(2) \\
                   & FY          & $-$8.478~~ & $-$7.760~~ & $-$28.50~~~ \\
 AV18pq/UIX        & VMC         & $-$8.22(1) & $-$7.47(1) & $-$27.21(2) \\
 AV8$^\prime$/UIX  & VMC         & $-$8.51(1) & $-$7.86(1) & $-$29.05(2) \\
                   & GFMC        & $-$8.68(1) & $-$8.03(1) & $-$29.82(2) \\
                                 &            &            &             \\
 AV18/TM$^\prime$(4.756)         & VMC & $-$8.26(1) & $-$7.50(1) & $-$27.51(2)\\
                                 & FY  & $-$8.444~~ & $-$7.728~~ & $-$28.36~~~\\
 AV18pq/TM$^\prime$(4.756)       & VMC & $-$8.22(1) & $-$7.46(1) & $-$27.42(2)\\
 AV8$^\prime$/TM$^\prime$(4.756) & VMC & $-$8.44(1) & $-$7.79(1) & $-$28.83(2)\\
                                 &            &            &             \\
 Nijm~I            & FY          & $-$7.736~~ & $-$7.085~~ & $-$24.98~~~ \\
 Nijm~II           & FY          & $-$7.654~~ & $-$7.012~~ & $-$24.56~~~ \\
 Nijm~I/TM(5.035)  & FY          & $-$8.392~~ & $-$7.720~~ & $-$28.60~~~ \\
 Nijm~II/TM(4.975) & FY          & $-$8.386~~ & $-$7.720~~ & $-$28.54~~~ \\
\end{tabular}
\label{tab:vmc}
\end{table}

\vfill

\end{document}